\documentclass[11pt]{article}
\usepackage{moriond,epsfig,pennames}

\def\Journal#1#2#3#4{{#1} {\bf #2}, #3 (#4)}

\def\NPB{{\em Nucl. Phys.} B}
\def\PLB{{\em Phys. Lett.}  B}

\def\be{\begin{equation}}
\def\ee{\end{equation}}
\def\bea{\begin{eqnarray}}
\def\eea{\end{eqnarray}}

\newcommand{\ipb}{\mbox{$\mathrm {pb}^{-1}$}}

\newcommand{\epem}{\mbox{$\mathrm{e}^+\mathrm{e}^-$}}
\newcommand{\WpWm}{\mbox{$\mathrm{W}^+\mathrm{W}^-$}}
\newcommand{\qqbar}{\mbox{$\mathrm{q\overline{q}}$}}
\newcommand{\GEVcc}{\mbox{$\mathrm{GeV}/{{\it c}^2}$}}
\newcommand{\MW}{\mbox{$m_{\mathrm{W}}$}}
\newcommand{\Vckm}[1]{\ensuremath{V_\mathrm{#1}}}
\def\ra{\mbox{$\rightarrow$}} 
\def\rs{\mbox{$\sqrt{s}$}} 

\begin{document}
\vspace*{4cm}
\title{W-PAIR PRODUCTION AT LEP}

\author{ P.AZZURRI }

\address{European Organization for Nuclear Research\\
CERN, EP Division, CH-1211 Gen\`eve 23, Switzerland.}

\maketitle\abstracts{
We review here the measurements of W-pair production and W decay rates 
performed by the four LEP experiments at \rs\ from 161 to 209~GeV, 
and their relevance in support of the standard electroweak model.
}

\section{Introduction}
From July 1996 to November 2000 the LEP collider at CERN has 
delivered \epem\ collisions above the kinematic threshold of the 
W-pair production, at \rs\ ranging from 161 to 209~GeV, for a total 
integrated luminosity of about 700~\ipb\ per experiment.
The total collected data samples yielded 
about 10000 W-pair events per experiment,
and allowed the first precision measurements of partial and total
WW cross sections at different \rs\ energy points, as well as the 
first direct determinations of all W decay Branching Ratios, 
to be achieved.

\subsection{W-pair production in the Standard Model}
\begin{figure}[htb]
  \vspace{-0.8cm}
  \centerline{\epsfig{file=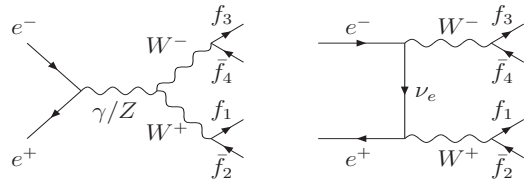,height=3.0cm,width=8cm}}
\caption{W-pair production in \epem collisions: 
  Standard Model CC03 diagrams.}
\label{fig:diaww}
\end{figure}

In the electroweak model the $\epem\ra\WpWm$ reaction is described,
at lowest level, by the three diagrams shown in Figure~\ref{fig:diaww}
and referred to as CC03 diagrams. 
Among these production diagrams the $t$-channel $\Pgne$-exchange
has well established We$\Pgne$ coupling vertices, while 
the two diagrams with the $s$-channel $\gamma$ and Z boson exchanges
involve WW$\gamma$ and WWZ triple gauge couplings 
that were not determined
before the start of the LEP data taking above the W-pair threshold.
The r\^ole of the triple gauge couplings is crucial. 
In fact for the W-pair production the $t$-channel 
contribution alone grows rapidly with energy 
($\sigma_\nu\sim s$), violating unitarity, while 
the addition of the interference effects of the $\gamma$ and 
Z $s$-channel cancels this divergent behaviour and leads to a 
cross-section which decreases at high energy ($\sigma\sim \ln s/s$).
The CC03 W-pair production is therefore crucially determined 
by cancellations arising from the coupling coefficients 
dictated by the electroweak model's SU(2)$\otimes$U(1) 
gauge invariance~\cite{sm}.

\section{Event selections}
The event topologies and selection criteria used
to select W-pair events~\cite{al,de,l3,op}
are mainly differentiated according to the possible 
leptonic (W$\ra\ell\nu$) or hadronic (W$\ra$qq')  
decay of each W boson. The resulting four decay fermions 
are energetic ($E_f\sim \rs/4 \sim$50~GeV) and mutually 
isolated.

In the case of fully leptonic decays both W's decay 
in a lepton-neutrino pair. This decay generates a low
particle multiplicity and a large missing momentum transverse to
the beam line. Further the two selected primary leptons  
will be energetic 
and have a large acoplanarity in the plane
transverse to the beam line. The achieved signal selection 
efficiency is 60-80\% with a purity of 90-80\%.
Residual backgrounds come from $\epem\ra\tau^+\tau^-$ 
and $\gamma\gamma\ra\ell^+\ell^-$ processes.

In the semi-leptonic channel (WW$\ra\ell\nu$qq'), the 
primary lepton is energetic and isolated from the two jets 
of the hadronic system. The primary neutrino will 
produce a large missing momentum vector, also isolated in space.
In this channel signal efficiencies are in the 60-90\% range with
corresponding purities of 95-85\%. Main backgrounds originate 
from $\qqbar$ events with energetic leptons from heavy quark 
decays, and other four fermion processes as 
$\epem\ra{\rm ZZ}^\ast, {\rm Zee}$.

Fully hadronic W-pair decays result in four jet events without
missing energy, characterised by a very large particle multiplicity 
with a spherical momentum distribution. Many observables
based on the global event topology and on the jets kinematics 
are used to select these events, resulting in signal efficiencies
of 60-80\% and purities of 85-75\%. The largely dominant 
background is generated from non-radiative \qqbar\ events 
with hard QCD gluon emissions.
\begin{figure}[htb]
  \centerline{\epsfig{file=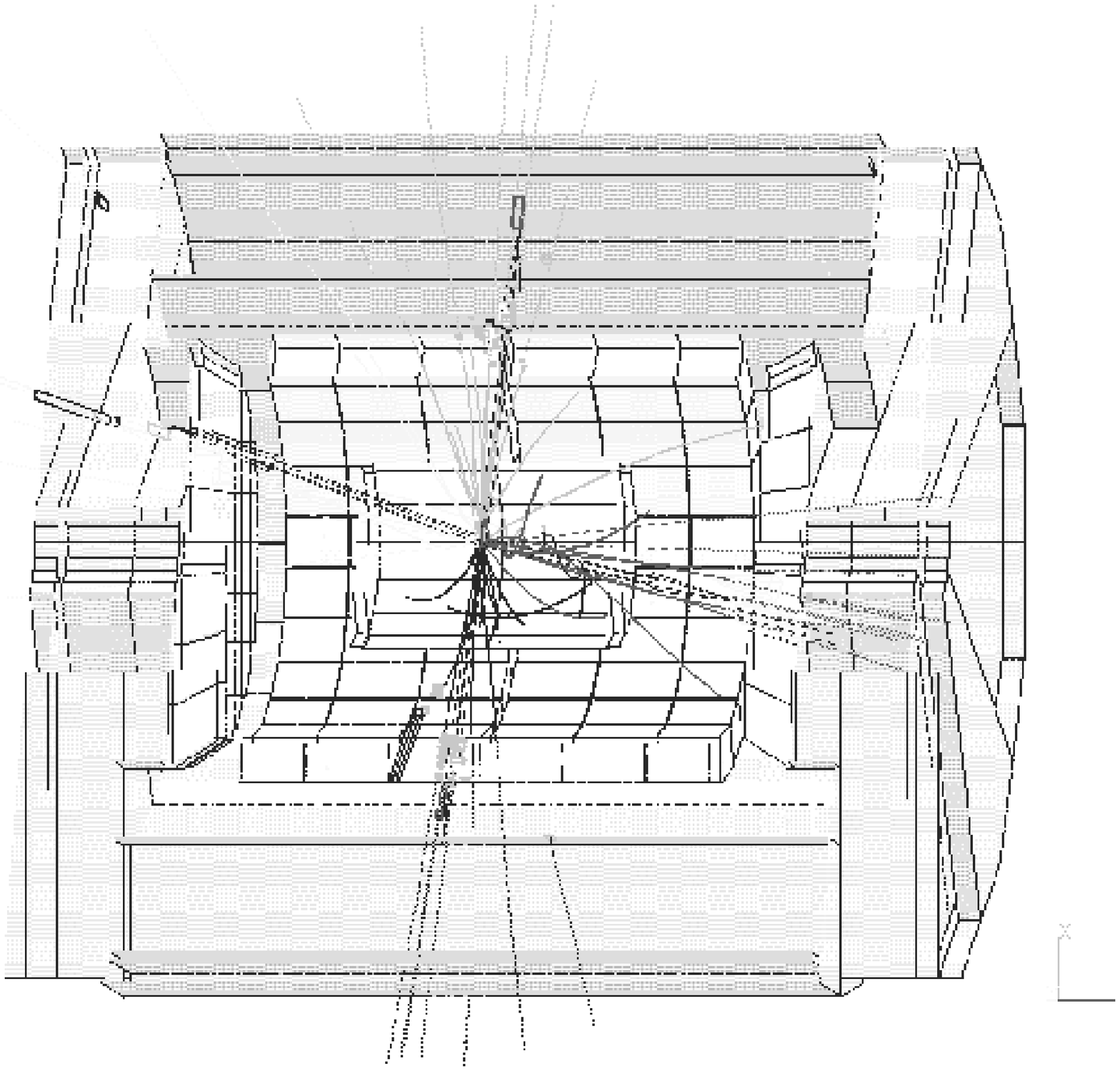,height=7.0cm}
    \hspace{1cm}\epsfig{file=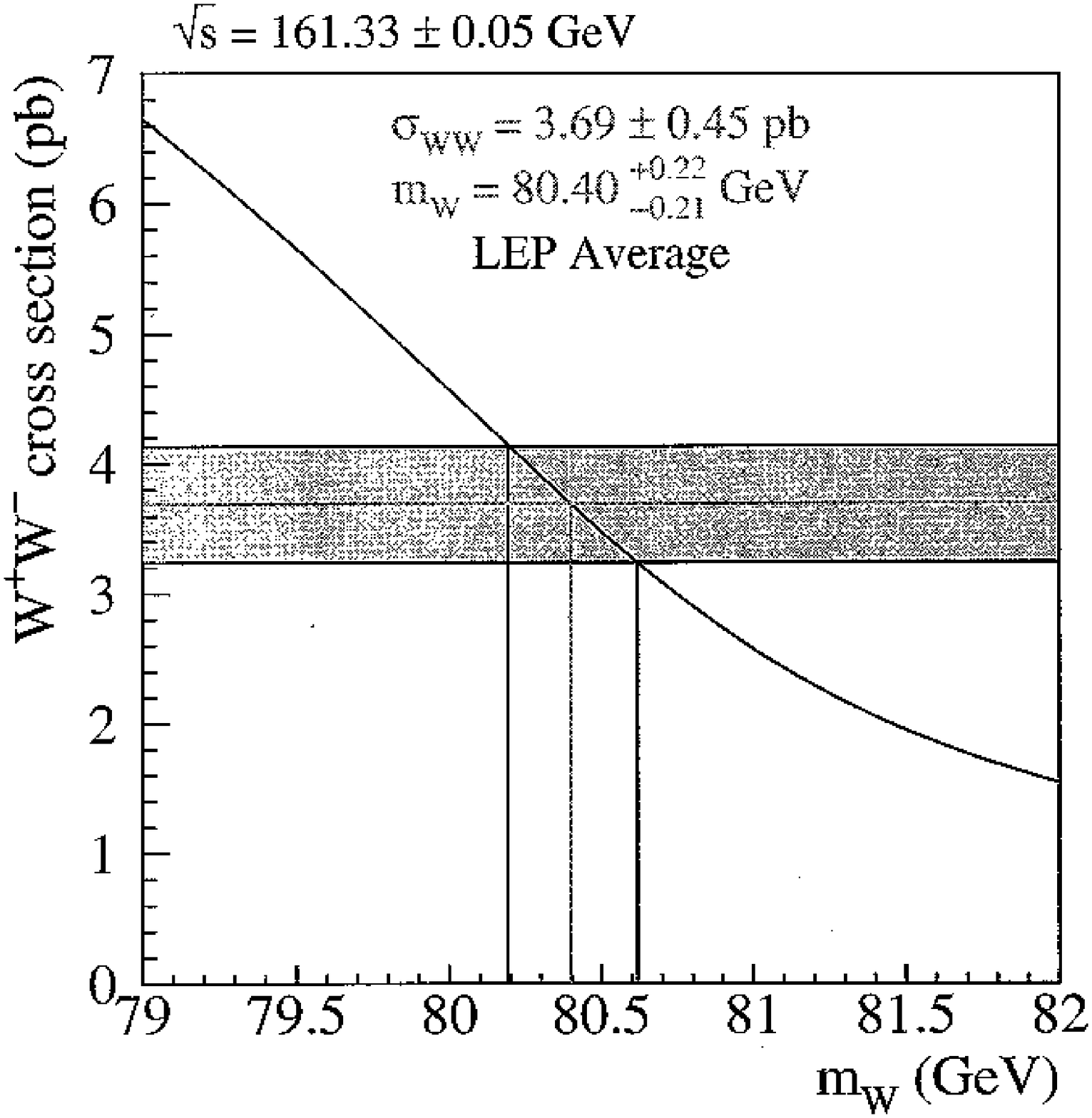,height=7.0cm}}
\caption{(left): The first W-pair event seen at LEP: a 
\qqbar\qqbar\ decay in DELPHI at \rs=161~GeV.
(right): W-pair cross section at \rs=161~GeV as a function of 
the W mass. The shaded band indicates the LEP determination.
\label{fig:tww}}
\end{figure}

\section{Extraction of W-pair cross sections}
\begin{figure}[htb] \vspace{-0.7cm}
  \centerline{\epsfig{file=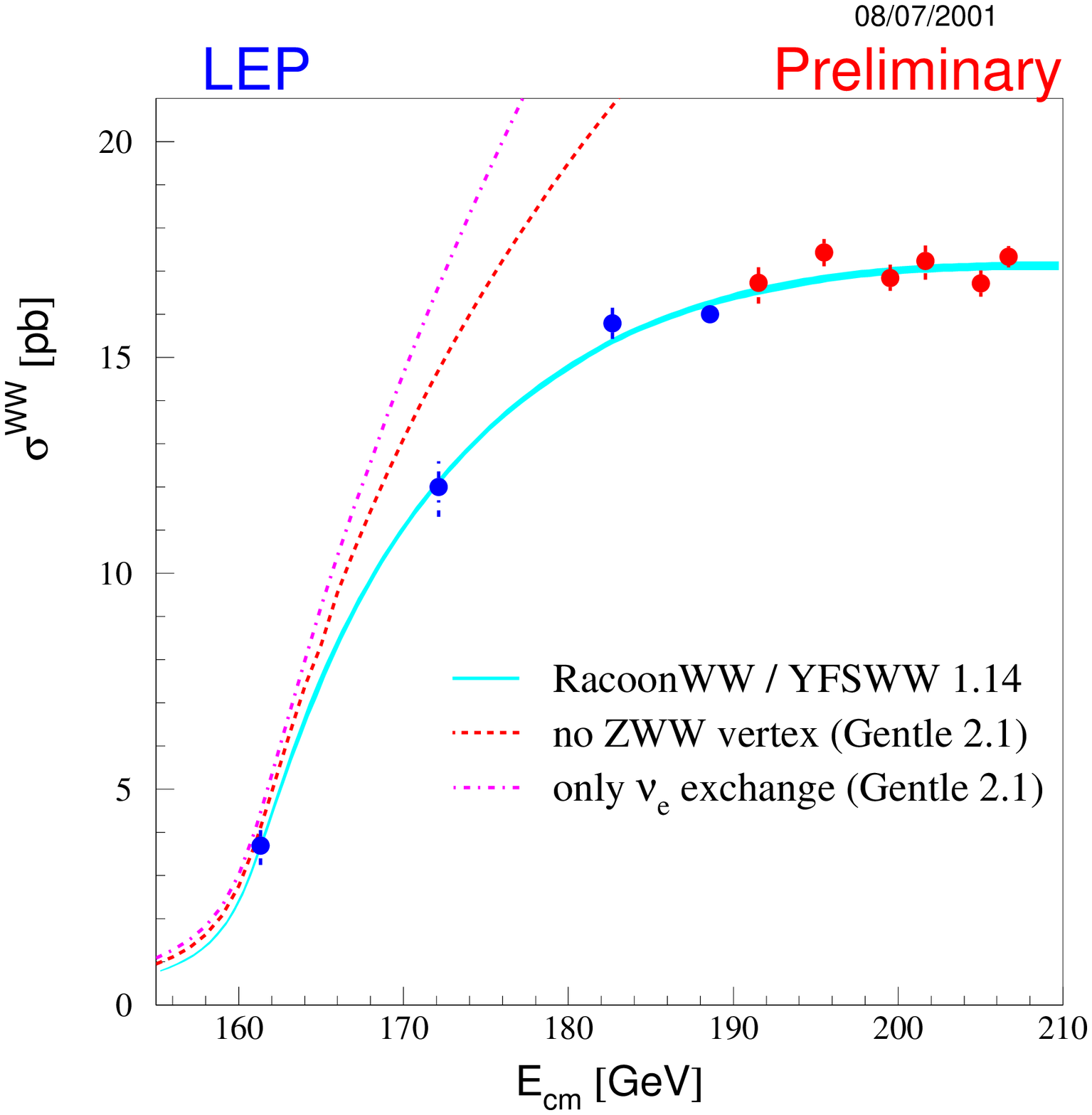,height=7.0cm}
    \hspace{1cm}\epsfig{file=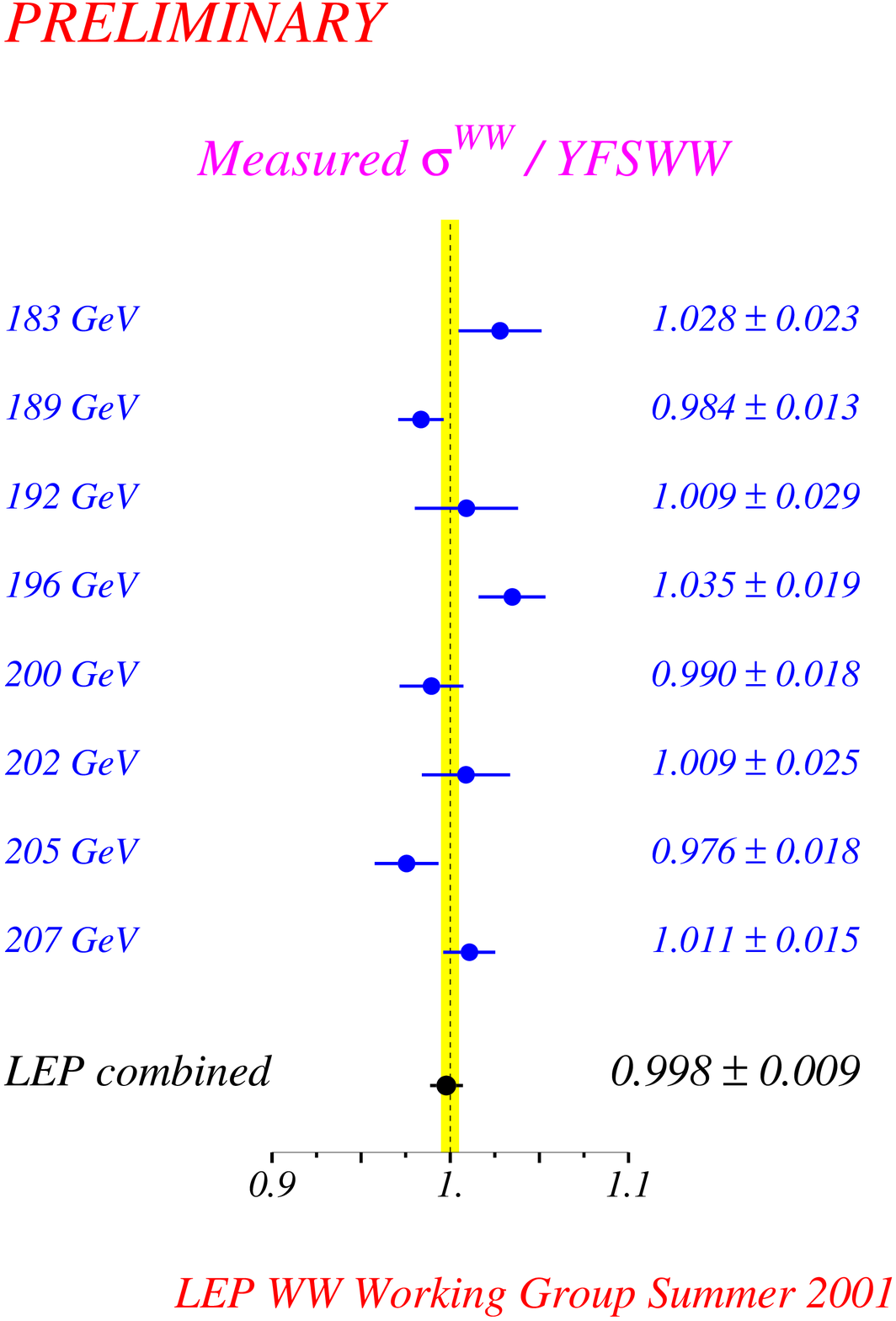,height=8.0cm}}
\caption{(left): Measurements of the total W-pair production 
cross section and comparisons with Standard Model predictions as a function 
of \rs . The data points clearly proof the existence of both the 
WW$\gamma$ and WWZ couplings with 
the expected electroweak SU(2)$\otimes$U(1) amplitudes.
(right): Ratio of measured W-pair cross sections to Standard Model 
calculations with O($\alpha$) radiative corrections. 
The overall precision on $\sigma^{\rm WW}$ 
is around 1\%.}
\label{fig:sww}
\end{figure}

Partial and total cross sections are determined through
maximum likelihood fits to the number of selected events 
in each decay channel, assuming Poisson distributions for 
the numbers of observed events.
The W-pair cross section yields~\cite{ewc} obtained with 
the full LEP data sample at \rs=161-207~GeV
are shown in figure~\ref{fig:sww}
and are compared with calculations
of the Standard Model CC03 production.
The agreement of experimental results with such  
theoretical expectations is the first clear evidence 
of the existence of both the WW$\gamma$ and WWZ couplings with 
the expected electroweak SU(2)$\otimes$U(1) coefficients.

It should be remarked that the theoretical calculations 
include first order electroweak radiative corrections 
in the so called Double Pole Approximation~\cite{dpa} (DPA), 
and that the data results disfavour calculations without
O($\alpha$) corrections by almost 2.5 sigmas. 
This result tells us that the overall LEP precision 
($\sim$1\%) obtained on $\sigma^{\rm WW}$ allowed us to test 
the Standard Model CC03 production at the one loop level.

\subsection{W mass from the threshold production rate}
As shown in figure~\ref{fig:tww}, the 
W-pair cross-section at the production kinematic threshold
is very sensitive to the W mass value, so that the LEP 
measurement of $\sigma^{\rm WW}$ at \rs=161~GeV 
has allowed the independent determination of 
$\MW= 80.40 ^{+0.22}_{-0.21}~\GEVcc$, to be achieved~\cite{thr}.

\section{Extraction of W decay Branching Ratios}
The W decay fractions into leptons and hadrons are 
extracted in a similar way as the cross-sections, 
by fitting the numbers of events selected in 
different decay channels. The LEP results are the 
first direct measurements of all W decays, 
and are summarized in table~\ref{tab:brw}.
The results of the leptonic branching ratios 
constitute the first test of the e-$\mu$-$\tau$
lepton universality of charged currents at the \MW\
scale. The two-by-two comparisons of the three leptonic fractions 
confirm this universality at the 2-3\% level.
The W hadronic branching ratio is extracted assuming
universal lepton flavour couplings, the result is  
in good agreement with Standard Model predictions and can be  
used to constrain the CKM quark mixing matrix $\Vckm{ij}$ 
yielding
$$
\sum_{i=u,c; j=d,s,b}\left|\Vckm{ij} \right|^2 = 2.039\pm0.025
\hspace{1cm}{\rm and}\hspace{1cm}
|\Vckm{cs}|= 0.996\pm0.013.
$$
These results constitute a good test of the CKM matrix unitarity in the 
first two families, and 
the best current constraint on the $|V_{cs}|$ amplitude.

\begin{table}[htb]
\caption{Summary of preliminary W branching 
fractions measurements obtained with the full LEP W-pair data samples.
The hadronic branching ratios are obtained under the 
assumption of lepton universality.\label{tab:brw}}
\vspace{0.4cm}
\begin{center}
\begin{tabular}{|lr|cccc|c|}
\hline
 & & ALEPH & DELPHI & L3 & OPAL & LEP \\ \hline
BR(W$\ra\Pe\Pgne$)   &[\%]& 10.95$\pm$0.31 & 10.36$\pm$0.34 & 
 10.40$\pm$0.30 & 10.40$\pm$0.35 & 10.54$\pm$0.17 \\
BR(W$\ra\Pgm\Pgngm$) &[\%]& 11.11$\pm$0.29 & 10.62$\pm$0.28 & 
  9.72$\pm$0.31 & 10.61$\pm$0.35 & 10.54$\pm$0.16 \\
BR(W$\ra\Pgt\Pgngt$) &[\%]& 10.57$\pm$0.38 & 10.99$\pm$0.47 & 
 11.78$\pm$0.43 & 11.18$\pm$0.48 & 11.09$\pm$0.22 \\ \hline
BR(W$\ra$hadrons)    &[\%]& 67.33$\pm$0.47 & 68.10$\pm$0.52 & 
 68.34$\pm$0.52 & 67.91$\pm$0.61 & 67.92$\pm$0.27 \\ \hline
\end{tabular}
\end{center}
\end{table}

\section{Conclusions}
The measurements of W production and decay rates at LEP
has been a nice and exciting work, providing diverse 
and crucial new results to particle physics  
on the structure of gauge boson couplings, on the W mass,
on lepton universality and on quark mixing.

\section*{Acknowledgements}
I would like to express my gratitude to the organisers of these
Rencontres for providing a stimulating and
friendly atmosphere and for giving me the opportunity to
present these results. My participation was partly supported
by the European Commission.

\end{document}